\documentclass[%
 aip,
 jmp,%
 amsmath,amssymb,
reprint,%
]{revtex4-1}

\usepackage{graphicx}
\usepackage{dcolumn}
\usepackage{bm}

\begin{document}


\title{Probabilistic inverse design for self assembling materials}

\author{R. B. Jadrich}
\author{B. A. Lindquist}
\author{T. M. Truskett} 
 \email{truskett@che.utexas.edu}
\affiliation{ 
McKetta Department of Chemical Engineering, University of Texas at Austin, Austin, Texas 78712, USA
}%

\date{\today}

\begin{abstract}
One emerging approach for the fabrication of complex architectures on the nanoscale is to utilize particles customized to intrinsically self-assemble into a desired structure. Inverse methods of statistical mechanics have proven particularly effective for the discovery of interparticle interactions suitable for this aim. Here we evaluate the generality and robustness of a recently introduced inverse design strategy [Lindquist et al., J. Chem. Phys. 145, 111101 (2016)] by applying this simulated-based, machine learning method to optimize for interparticle interactions that self-assemble particles into a variety of complex microstructures: cluster fluids, porous mesophases, and crystalline lattices. Using the method, we discover isotropic pair interactions that lead to self-assembly of each of the desired morphologies, including several types of potentials that were not previously understood to be capable of stabilizing such systems. One such pair potential led to assembly of the highly asymmetric truncated trihexagonal lattice and another produced a fluid containing spherical voids, or pores, of designed size via purely repulsive interactions. Through these examples, we demonstrate several advantages inherent to this particular design approach including the use of a parametrized functional form for the optimized interparticle interactions, the ability to constrain the range of said parameters, and compatibility of the inverse design strategy with a variety of simulation protocols (e.g., positional restraints). 
%
\end{abstract}

\maketitle

\section{Introduction} 
The engineering of materials with nanoscale structural features is an increasingly relevant design challenge, the difficulty of which grows rapidly with diminishing structural feature size. One emerging strategy in this area is to tailor interparticle interactions between material building blocks to drive their spontaneously assembly into specified architectures. Working as either a stand alone technique or in tandem with other small scale patterning technologies, like lithography, novel nanostructured morphologies can be realized by self assembly. Colloids are particularly amenable to such an approach due to the wide variety of experimentally tunable interparticle interactions that are possible, including (approximately) isotropic pair interactions for nearly spherical colloids that depend on factors such as the material composition of the particle and the ligand coating, as well as the quality and composition of the solvent.~\cite{CollInt1,CollInt2} Directionally specific, short-ranged interactions between ``patchy'' particles or excluded volume interactions between particles of different shapes are possible as well.~\cite{patchy_review1,patchy_review2} Furthermore, various classes of interactions can be controlled in an approximately orthogonal fashion, giving rise to a rich design space for assembly. From colloidal building blocks, numerous self-assembled structures have been observed experimentally. The two-dimensional Kagome lattice was assembled via triblock Janus particles, for instance.~\cite{kagome} Other self-assembled microstructures include tubes, micelles, lamellar sheets and viral capsid-like structures, to mention a few.~\cite{capsid,BCPs}

The sheer variety of interactions available for colloids presents an intriguing challenge. How does one rationally select appropriate interactions to drive assembly of a desired structural motif or target a certain property or functionality? For some relatively simple cases, intuition or an empirical method may suffice. However, more complex architectures or non-trivial colloidal interactions necessitate a more systematic approach. In recent years, many inverse design (ID) strategies have emerged to allow for the discovery of interactions that maximize a selected figure of merit in an (ideally) automated fashion.~\cite{property_id,template_assist_id,ST_inv_des_review,AJ_inv_des_review,ID_crystals_TS,ID_crystals,swarm_id} In addition to ID algorithm development, advances in computational power have enabled ID schemes to play an increasingly influential role in the rational design of materials.

\begin{figure}
  \includegraphics[width=3.37in,keepaspectratio]{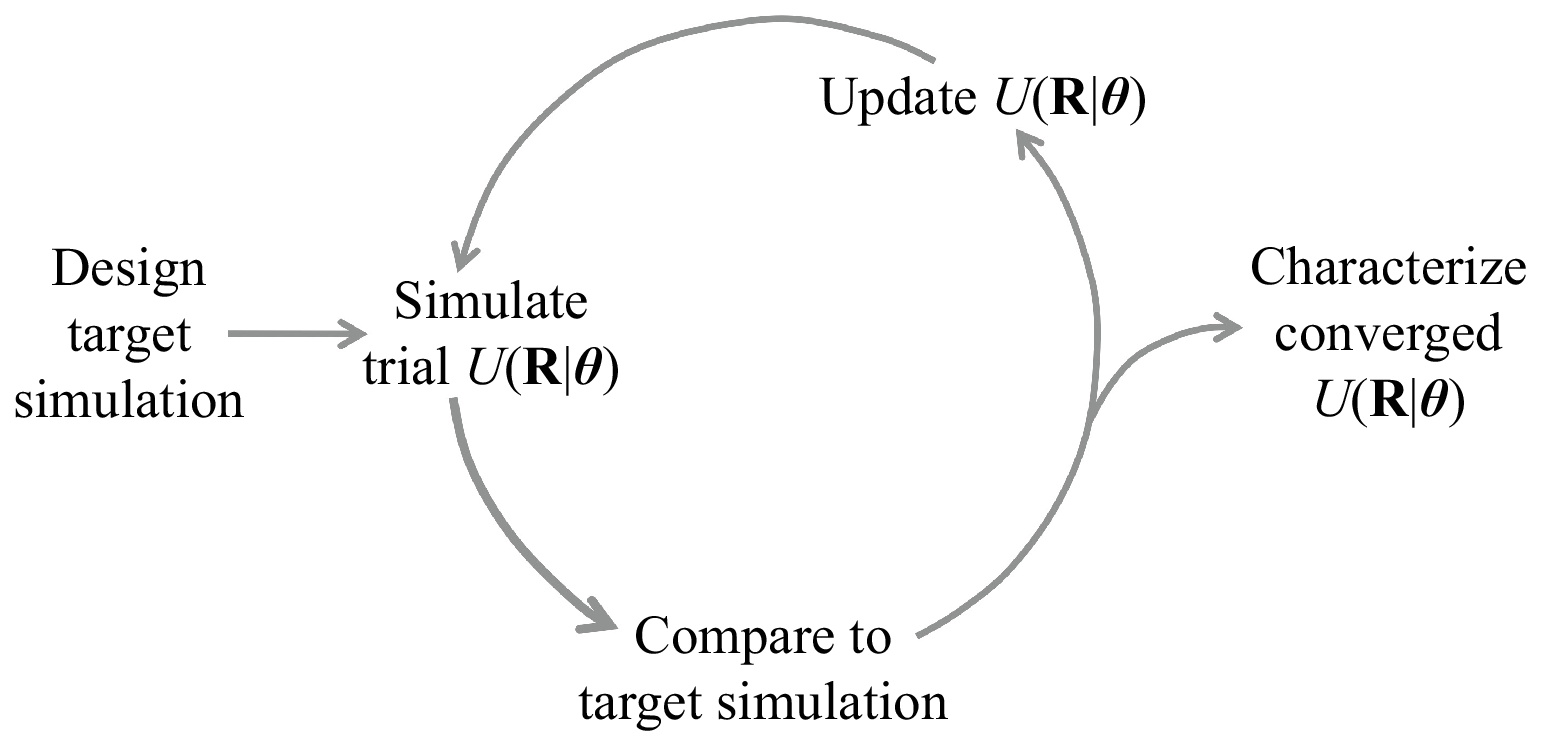}
  \caption{Schematic representation of an iterative Inverse Design (ID) procedure, where $U(\textbf{R}|\boldsymbol{\theta})$ indicates any general interaction potential for a given particle configuration, $\textbf{R}$.}
  \label{fgr:IDcycle}
\end{figure}

In general, ID relies on an optimization toward a well-defined and measurable figure of merit. For self-assembly problems, figures of merit based on structural data are a natural choice, and several ID strategies focused on structure have been developed within the field of mesoscale modeling. For instance, biomolecular or polymeric coarse-graining techniques aim to reproduce desired structural signatures of an all-atom simulation using a reduced dimensionality model, typically where a single bead represents multiple atoms.~\cite{IBI1,IBI2,IBI3,test_of_ID_schemes} We refer to the desired structures (e.g., from the all-atom description in the prior case) to be emulated by the ID interaction potentials as target configurations. For application of this framework to self-assembly problems, we generate an ensemble of target configurations via a simulation where we apply (often complex and unphysical) many-body constraints to the particles in order to ensure that the desired structural motifs are present. We then systematically determine the ID interactions, i.e., the simplified interactions that best reproduce the structure of the target configurations according to the figure of merit. The optimization used for this purpose proceeds in an iterative fashion, where the difference between the data derived from the target simulation and those from the ID interaction simulation informs the update to the potential to be used in the next iteration. The scheme proceeds until either satisfactory convergence is achieved or the desired property is realized. The overall strategy is summarized in Fig.~\ref{fgr:IDcycle}.

The techniques that have been developed in the coarse-graining community for discovering pair interactions that will best reproduce the target structure are typically aimed at matching the probability distributions of configurations in the target and ID simulations~\cite{RE_ID,general_RE,IBI1} or the forces on the particles in those two simulations.~\cite{voth_ID,IBI1} While such methods have enjoyed some use in sophisticated coarse-graining applications, they have not been typically applied to self-assembly problems. In this work, we will primarily minimize the relative entropy (RE), equivalent to the well-known maximum likelihood approach in probability and statistics, in order to carry out our ID calculations.~\cite{RE_ID,general_RE} There are many advantages of the RE approach, including a rigorous mathematical framework for the method, well-behaved convergence of the update scheme, and the ease of employing constrained functional forms for the ID interactions. This last point is particularly relevant to colloids, given the many simplified physics-based functional forms that exist for computational modeling of such interparticle interactions.~\cite{CollInt1,CollInt2} 
More generally though, constrained functional forms allow one to (1) limit the number of features, (2) impose bounds on parameters, (3) implement a well-defined and smooth cutoff, and (4) easily optimize even if the features of the radial distribution function are much longer-ranged than the interaction.

Of the many possible tunable parameters available, the focus of this paper is on the discovery of isotropic pair interactions that self-assemble into interesting structures, though the maximum likelihood formalism is not restricted to such cases. There are several benefits to discovering pair interactions for a particular self-assembly problem. First, optimization of pair interactions is more computationally tractable than more complex forms. If a pair interaction exists that self assembles into a desired structure, it can be taken as proof of realizability with more complex (3-body or higher) interactions as well. Moreover, pair interactions can provide high-level, interpretable insights into the physics necessary to achieve a particular structure. Relevant to the present work, it has been established that isotropic pair potentials, particularly those possessing competing attractions and repulsions, can self-assemble into a variety of microphase-separated states, such as clusters, columns, and lamellae.~\cite{mod_phases1,mod_phases2,mod_phases3,mod_phases4,mod_phases5,SALR2,SALR3,pores,pores2,simulated_phases,postulated_phases_sear,mean_field_assembly_0,mean_field_assembly_1,mean_field_assembly_2} Moreover, open lattices, such as diamond in three dimensions and honeycomb in two dimensions, have been assembled from \emph{purely repulsive} isotropic pair potentials.~\cite{BL_RE,ST_inv_des_review,AJ_inv_des_review,ID_crystals_TS,ID_crystals} The results presented herein not only demonstrate the utility of ID towards self-assembly problems, but also help to further explore the limits of assembly via pair interactions.  

In this Article, we first describe a generally applicable ID methodology for particle self-assembly. The approach is technically simple to implement, particularly for pair interactions. Beginning from a maximum likelihood formulation, we provide an overview of the update scheme, as well as provide additional computational details in Sect.~\ref{sec:methods}. In Sect.~\ref{sec:clusters}-\ref{sec:crystals}, we present ID results (in conjunction with relevant background material) for several classes of particle assemblies: clusters, pores, and crystalline lattices, respectively. Some prior results are recalled for context, but in large the results presented here are new and address various proposed extensions of the prior work encompassing (1) result generality, (2) extensions to different interaction classes, and (3) the ability of the scheme to handle increasing target complexity. Finally in Sect.~\ref{sec:conclusions}, we conclude and comment on various extensions of RE minimization towards more complex design goals.

\section{Computational Considerations}
\label{sec:methods}

The location of each particle $i$ is described by a $D$-dimensional vector $\textbf{r}_i$, and a full configuration of $N$ particles comprises one such vector, $\textbf{R}\equiv[\textbf{r}_{1},\textbf{r}_{2},...,\textbf{r}_{N}]$. 
As discussed in the Introduction, a suitable optimization target is required for ID. To this end, creating a target ensemble entails formulating a target probability distribution, $P_{\text{tgt}}(\textbf{R})$, from which an ensemble of particle configurations 
can be sampled that embody the desired microstructural organization. In this work, we employ target models motivated by statistical mechanics: $P_{\text{tgt}}(\textbf{R})$ is a Boltzmann distribution governed by many-body potentials or external fields with some arbitrary notion of a ``temperature''. While the interactions used to produce the target distributions are themselves physically unrealistic, they are tailored such that the desired organizational motifs dominate the target ensemble. More details on the constraints employed in each individual case are provided in Sect.~\ref{sec:clusters}-\ref{sec:crystals}. 

\subsection{Probabilistic machine learning}

For systematic optimization of the dimensionless tunable interaction parameters $\boldsymbol{\theta}$ of the ID potential, we exploit maximum-likelihood model fitting, a strategy commonly adopted in probabilistic machine learning and formally equivalent to the relative entropy approach of Shell and coworkers.~\cite{RE_ID,general_RE} We provide a brief review of the fundamental principles behind the method below. 

The role of maximum-likelihood is to maximize the probability (or equivalently, the log probability) that a classical, particle-based system at thermal equilibrium will sample the same configurations generated by $P_{\text{tgt}}(\textbf{R})$. Here we consider the case of constant particle number ($N$), volume ($V$), and temperature ($T$), whereby the probability~\footnote{This quantity is actually a probability distribution, though we use the terminology interchangeably without consequence.} for realizing any single configuration, $\textbf{R}$, conditioned on $\boldsymbol{\theta}$, is simply the Boltzmann weight
\begin{equation} \label{eqn:canonical_prob}
P(\textbf{R}|\boldsymbol{\theta})\equiv \dfrac{\exp[-\beta U(\textbf{R}|\boldsymbol{\theta})]}{Z(\boldsymbol{\theta})}
\end{equation} 
where $U(\textbf{R}|\boldsymbol{\theta})$ is the tunable potential energy for configuration $\textbf{R}$, and
\begin{equation} \label{eqn:part_func}
Z(\boldsymbol{\theta})\equiv\int d\textbf{R} \exp[-\beta U(\textbf{R}|\boldsymbol{\theta})]
\end{equation}
is the configuration partition function. With these definitions, the the quantity we seek to maximize is~\cite{BL_RE}
\begin{equation} \label{eqn:log_likelihood_2}
\langle\text{ln}P(\textbf{R}|\boldsymbol{\theta})\rangle_{P_{\text{tgt}}(\textbf{R})} = -\langle\beta U(\textbf{R}|\boldsymbol{\theta})\rangle_{P_{\text{tgt}}(\textbf{R})} - \text{ln}Z(\boldsymbol{\theta})
\end{equation}
Eqn.~\ref{eqn:log_likelihood_2} is effectively a measure of the ``overlap'' of $P(\textbf{R}|\boldsymbol{\theta})$ with $P_{\text{tgt}}(\textbf{R})$. In fact, maximizing Eqn.~\ref{eqn:log_likelihood_2} is equivalent to minimizing the Kullback-Leibler (KL) divergence--an information theory measure of the ``distance'' between two distributions.~\cite{KL_div} In the statistical mechanical coarse-graining literature, the RE approach is often formulated in terms of minimizing the KL divergence with respect to $\boldsymbol{\theta}$~\cite{RE_ID,IBI1} and is therefore equivalent to maximum-likelihood learning.

In order to carry out the maximization, we employ a gradient ascent optimization algorithm, 
\begin{equation} \label{eqn:grad_descent}
\boldsymbol{\theta}^{(k+1)} = \boldsymbol{\theta}^{(k)} + \alpha \big[\boldsymbol{\nabla}_{\boldsymbol{\theta}} \langle\text{ln}P(\textbf{R}|\boldsymbol{\theta})\rangle_{P_{\text{tgt}}(\textbf{R})}\big]_{\boldsymbol{\theta}=\boldsymbol{\theta}^{(k)}}
\end{equation}
where $k$ indexes the iteration step, and $\alpha$ is the empirically determined learning rate chosen to yield a stable optimization. Utilizing Eqn.~\ref{eqn:log_likelihood_2} along with the definition of the canonical probability distribution (Eqn.~\ref{eqn:canonical_prob}) and the partition function (Eqn.~\ref{eqn:part_func}), we find that 
\begin{equation} \label{eqn:grad_log_likelihood}
\begin{split}
& \boldsymbol{\nabla}_{\boldsymbol{\theta}} \langle\text{ln}P(\textbf{R}|\boldsymbol{\theta})\rangle_{P_{\text{tgt}}(\textbf{R})} \\
& = \langle \boldsymbol{\nabla}_{\boldsymbol{\theta}} \beta U(\textbf{R}|\boldsymbol{\theta})\rangle_{P(\textbf{R}|\boldsymbol{\theta})} -\langle \boldsymbol{\nabla}_{\boldsymbol{\theta}} \beta U(\textbf{R}|\boldsymbol{\theta})\rangle_{P_{\text{tgt}}(\textbf{R})} 
\end{split}
\end{equation}
Iterations are stopped once a sufficiently small gradient has been achieved.

\subsection{Specific strategy for pair interactions}

In this paper, we focus exclusively on the case of single component systems with isotropic pair interactions,
\begin{equation} \label{eqn:pair_potential}
U(\textbf{R}|\boldsymbol{\theta})\equiv\sum_{i=1}^{N}\sum_{j=i+1}^{N}u(r_{i,j}|\boldsymbol{\theta})
\end{equation}
where $r_{i,j}=|\textbf{r}_{i}-\textbf{r}_{j}|$ is the distance between particles $i$ and $j$. As a consequence, only the isotropic pair correlations contained in $P(\textbf{R}|\boldsymbol{\theta})$ or $P_{\text{tgt}}(\textbf{R})$ are required to compute the averages in Eqn.~\ref{eqn:grad_log_likelihood}; in general, when the averaged quantity depends on fewer degrees of freedom than supported by the the full probability distribution, only the reduced (marginalized) probability distribution over those degrees of freedom is required to perform the average.


Therefore, plugging Eqn.~\ref{eqn:pair_potential} into Eqn.~\ref{eqn:grad_log_likelihood} and summing identical terms in the large $N$ limit yields
\begin{equation} \label{eqn:grad_log_likelihood_pair}
\begin{split}
& \boldsymbol{\nabla}_{\boldsymbol{\theta}} \langle\text{ln}P(\textbf{R}|\boldsymbol{\theta})\rangle_{P_{\text{tgt}}(\textbf{R})} \\
& = \dfrac{N^{2}}{2}\Big[\langle \boldsymbol{\nabla}_{\boldsymbol{\theta}} \beta u(r|\boldsymbol{\theta})\rangle_{P(r|\boldsymbol{\theta})} - \langle \boldsymbol{\nabla}_{\boldsymbol{\theta}} \beta u(r|\boldsymbol{\theta})\rangle_{P_{\text{tgt}}(r)}\Big]
\end{split}
\end{equation}
where the $P(r)$ are the isotropic two-particle probability distributions quantifying the probability of a pair of particles to be a radial distance $r$ apart. Formally, the $P(r)$ are obtained by integrating the full $P(\boldsymbol{R})$ over the $N-2$ excess particle degrees of freedom as well as the orientation degrees of freedom between the remaining particle pair. Since all particles are identical, any chosen pair of particles yields the same $P(r)$. These reduced probability distributions can be defined in terms of radial distribution functions [$g(r)$] as $P(r)\equiv s(D)r^{D-1}g(r)/V$ where $s(D)$ is the surface area of a unit radius $D$-dimensional sphere yielding 
\begin{equation} \label{eqn:grad_log_likelihood_pair_2}
\begin{split}
& \boldsymbol{\nabla}_{\boldsymbol{\theta}} \langle\text{ln}P(\textbf{R}|\boldsymbol{\theta})\rangle_{P_{\text{tgt}}(\textbf{R})} \\
& = \dfrac{s(D)}{2} \rho N   \int_{0}^{\infty} drr^{D-1}[g(r|\boldsymbol{\theta})-g_{\text{tgt}}(r)]\boldsymbol{\nabla}_{\boldsymbol{\theta}}\beta u(r|\boldsymbol{\theta})
\end{split}
\end{equation}
where [$g(r|\boldsymbol{\theta})$] is the optimized $g(r)$, [$g_{\text{tgt}}(r)$] is the target $g(r)$, and $\rho$ is the number density. Thus, the final iterative update framework for \emph{pair interactions} can be succinctly stated as 
\begin{equation} \label{eqn:pair_update}
\begin{split}
& \boldsymbol{\theta}^{(k+1)}=\boldsymbol{\theta}^{(k)} \\
& +\alpha \int_{0}^{\infty} dr \dfrac{r^{D-1}}{\sigma^D} \big[g(r| \boldsymbol{\theta}^{(k)})-g_{\text{tgt}}(r)\big] \big[\boldsymbol{\nabla}_{\boldsymbol{\theta}} u(r|\boldsymbol{\theta})\big]_{\boldsymbol{\theta}=\boldsymbol{\theta}^{(k)}} 
\end{split}
\end{equation}
where $\sigma$ is a length scale that renders the right hand side, in particular $\alpha$, dimensionless. 

In the limit of optimizing a potential that has been infinitesimally discretized over a finite range in $r$, the zero-point solution to Eqn.~\ref{eqn:grad_log_likelihood_pair_2} is that where $g(r|\boldsymbol{\theta})=g_{\text{tgt}}(r)$ for all $r$ in the relevant range. Therefore, any other iterative update scheme that converges to an exact matching in $g(r)$ is equivalent to RE minimization under the aforementioned conditions. Iterative Boltzmann Inversion (IBI) is one such heuristic approach that is commonly used in the literature on statistical mechanical coarse-graining.~\cite{IBI1,IBI2,IBI3} The associated update scheme is:
\begin{equation} \label{eqn:ibi_equation}
u^{(i+1)}(r)\equiv u^{(i)}(r)+\alpha k_{B}T \textnormal{ln}\bigg[\dfrac{g^{(i)}(r)}{g_{\text{tgt}}(r)}\bigg]
\end{equation}
where \(\alpha\) is again a tunable parameter to control the magnitude of the update. IBI can be technically simpler to implement than unconstrained RE calculations, and therefore, we often perform IBI calculations first to inform our choice for potential form in the (constrained) RE optimizations.

\subsection{Simulation and analysis}


All molecular dynamics (MD) simulations were performed in the canonical ensemble using the GROMACS molecular dynamics package,~\cite{GROMACS_1,GROMACS_2} version 4.6.5 or 5.0.6 using a time step of $dt/\sqrt{\sigma^{2}m\beta}\approx0.001$ ($m$ is the particle mass, $\beta\equiv k_{B}T$ where $k_{B}$ is Boltzmann's constant, and $T$ is the temperature) with leapfrog integration. A velocity-rescaling thermostat was used to control $T$ with a characteristic time constant $\tau=100dt$; the choice of $T$ is arbitrary because the optimization result is the reduced pair interaction, $\beta u(r|\boldsymbol{\theta})$ which is independent of optimization temperature. Periodic boundary conditions were employed in three dimensions for all simulations, except for the two-dimensional lattices, where boundaries were periodic in the x- and y-directions only. IBI calculations were performed via the Versatile Object-oriented Toolkit for Coarse-graining Applications (VOTCA) package,~\cite{VOTCA_1,VOTCA_2} through its GROMACS interface. Initial configurations corresponded to homogeneous fluids, and initial guesses for the pair interaction were generally weakly interacting; see Section A1 in the Appendix for one example. As expected for an equilibrium statistical mechanics-based approach, the optimization is insensitive to the precise initialization details. Simulation visualizations were created with Visual Molecular Dynamics (VMD).~\cite{VMD}. More specific details regarding each simulation--such as particle number and the use of positional restraints--are given in the Appendix.

In order to evaluate the success of the ID calculation, we analyze the simulations for fidelity to the targeted microstructure. In some cases, particularly for the lattice assemblies, this can be done by visual inspection. However, to quantify the size of microphase-segregated objects (i.e., clusters and pores), we calculate probability distribution functions to characterize these features. For analyzing cluster size, we adopt a simple interparticle distance definition of neighboring particles and summarize the statistics in the form of cluster-size distributions (CSDs). Specifically, two particles are considered ``connected'' if their center-to-center distance does not exceed a cutoff, $r_{\text{cut}}$, or if they are joined via a continuous pathway of neighboring particles.~\cite{CSD_1,CSD_2} In keeping with prior work,~\cite{ID_clusters,pores} we set $r_{\text{cut}} = 1.25\sigma$ in Sect.~\ref{sec:clusters} and $r_{\text{cut}} = 1.33\sigma$ in Sect.~\ref{sec:pores}, where $\sigma$ is the particle diameter.  The CSD [$P(n)$] is normalized such that it provides the probability that a randomly selected cluster comprises $n$ particles. The presence of a \emph{local maxima} in $P(n)$ at $n^{*}$ occurring in the range $1 \ll n^{*} \ll N$ indicates a preferred aggregate size and thus the presence of clusters.

For the porous architectures, we first randomly insert test spheres with a diameter of $2\sigma$ into each configuration, keeping those that do not overlap with the particles of the configuration (they can overlap with themselves). The inserted test spheres provide a definition of the porous space in the configurations. Discrete void regions are identified by performing a cluster analysis on the test spheres as described above, where overlapping test spheres are taken to be neighbors. The volume of each pore (union volume of the test spheres) is computed via Monte Carlo integration, and the pore size distribution (PSD) is characterized as a function of pore diameter after the pore volumes are equated to the diameter of an equi-volume sphere.

\section{Clusters}
\label{sec:clusters}

\subsection{Target Simulation}
Target configurations of fluids of ``ideal'' clusters (i.e., fluid-like particle aggregates that are roughly spherical in shape and monodisperse in size) were generated via Monte Carlo simulations of hard-core-like particles that are each assigned to clusters with a specified aggregation number and interact via a many-body potential as described in Ref.~\citenum{ID_clusters}. In brief, the interaction has two components that stabilize clusters. The first is an energy penalty for each cluster that is quadratic in the deviation of its radius of gyration from the specified size, chosen to be sufficiently strong so as to generate well-defined, spherical clusters, but weak enough to maintain liquid-like, intracluster particle structure and mobility. The second component is an isotropic Yukawa repulsion between the centers of mass of any two clusters which effectively enforces a minimum distance between any two particles in separate clusters. Empirically, it was found that larger clusters in the target ensemble require larger minimum separations between particles in adjacent clusters to stabilize the inverse design optimization.~\cite{ID_clusters} In the present study, we focus on clusters that each contain 32 particles, the largest equilibrium aggregates investigated in our prior work.~\cite{ID_clusters}

\subsection{Prior work}
The short-ranged attractive, long-ranged repulsive (SALR) isotropic pair interaction type is frequently used in simulation studies of equilibrium cluster fluids--of broad interest due to both their interesting microstructures and the potential implications of clusters for bulk material properties, such as viscosity.~\cite{jon1,jon2,ryanPRE,clusters1,clusters2,clusters3,many_cluster_potentials} This class of pair potentials is believed to provide a reasonable description of charge-stabilized colloids with short-range (e.g., osmotic depletion or hydrophobic) attractions. Typical ranges for the attractive interaction are 1-10\% of the particle diameter. The longer-ranged repulsive component of the potential is generally weak in amplitude and slowly varying, often represented as a (repulsive) Yukawa form. 

Instead of performing a tedious forward search for desired clustered morphologies, we utilized ID to discover a class of isotropic pair interactions that assemble particles into highly idealized clusters of a prescribed size that (1) were monodisperse and spherical, (2) had marked spatial and temporal definition, and (3) possessed intra- and inter-cluster liquid-like mobility.~\cite{ID_clusters} The resultant pair interactions (discovered via IBI) deviated from the traditional SALR model, possessing a broad attractive well with a narrow repulsive hump, as demonstrated in Fig.~\ref{fgr:FigureClusters}a. We found that the location of the repulsive barrier is directly related to the cluster size, occurring at sufficiently large $r$ that the barriers do not overlap with the other particles in the cluster, but rather collectively furnish a repulsive corona around it, analogous to the protective intercluster repulsion that was included in the target simulation.~\cite{ID_clusters} Because (1) the interactions are isotropic and (2) the position of the repulsive barrier and the cluster size are coupled, 
there is a direct relationship between the cluster diameter and the thickness of the resulting repulsive corona. This relationship explains why target simulations for bigger clusters required larger minimum separations between particles in different clusters to stabilize the ID scheme. 

\subsection{Results}
\label{subsec:cluster_results}
From our prior work, several open questions remain. First, the IBI potential contains many secondary features in addition to the primary attractive well and repulsive barrier. We previously postulated that these features were not essential to equilibrium cluster fluid behavior.  In the present work, we test this assumption by carrying out optimizations with constrained pair potential forms via RE minimization. Second, in prior work, we compared the clusters generated by our ID potential to those assembled with a more typical SALR model with a shorter-ranged attraction ($\sim0.01\sigma$). We found that the latter clusters were less ``ideal''. However, the SALR potential was chosen for the comparison via a limited forward search through parameter space, not via a targeted ID optimization, and therefore it does not likely represent the optimal SALR form for ideal clustering behavior. To remedy this shortcoming here, we constrain the width of the attractive well in the RE calculation to yield an optimized potential that more closely resembles a prototypical SALR interaction.  

\begin{figure}
  \includegraphics{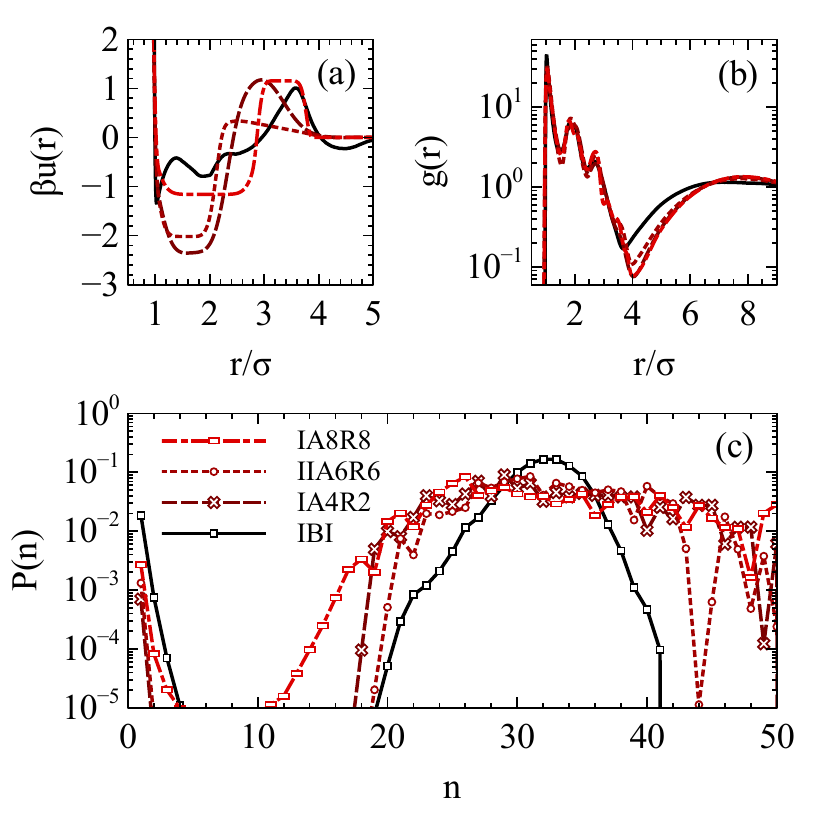}
  \caption{(a) Dimensionless pair interactions, $\beta u(r)$ discovered via ID (IBI or RE) for equilibrium cluster fluids and their associated (b) radial distribution functions, $g(r)$, and (c) cluster size distributions $P(n)$. The naming conventions and details of potentials obtained via RE (i.e., IA8R8, IIA6R6, and IA4R2) are provided in the Appendix.}
  \label{fgr:FigureClusters}
\end{figure}

To test whether particles interacting via a single attractive well and repulsive barrier can form an equilibrium fluid of ideal clusters, we used RE minimization to design potentials with appropriately constrained functional forms of this type (IA4R2, IIA6R6, and IA8R8), the specifics of which--and naming convention for--are provided in the Appendix.
The potentials mimic a square-well followed by a square-shoulder at larger $r$; larger numbers in the name indicate more square-like features for the attraction (A) and the repulsion (R), respectively. The roman numeral indicates the presence (II) or absence (I) of a linear ramp term in the interaction, which lends an asymmetry to the repulsive barrier.
Free to vary the range and amplitude of the attractive and repulsive components of the pair potential, the optimization finds interactions which exhibit a broad attractive well in concert with a significant repulsive hump (see Fig.~\ref{fgr:FigureClusters}a). Simulations verify that particles interacting via the resulting pair potentials form clusters with good center-of-mass mobility and internal fluidity.  
While the constraints imposed by the functional forms employed in the RE calculation necessitate degraded matching between the target and optimized $g(r)$ relative to IBI, the $g(r)$ for each system of particles (Fig.~\ref{fgr:FigureClusters}b) shows the characteristic structural features of a strongly clustered fluid:  a broad liquid droplet regime followed by an interfacial depletion zone and then low-amplitude oscillatory cluster-cluster correlations. All are similar to the $g(r)$ of cluster fluids of potentials designed from IBI--which nearly perfectly reproduces $g_{\text{tgt}}(r)$--but deviations, particularly in the interfacial depletion zone, are evident. 

Analysis of the CSD for each system (Fig.~\ref{fgr:FigureClusters}c) indicates that the more restricted interactions were able to generally reproduce the desired microstructure of a clustered fluid with aggregates of order 30 particles. Each potential yields crisp monomer-cluster separation and closely achieves the desired size, with average cluster sizes of $\langle n \rangle=32.8$, 33.3 and 32.9 for the IA4R2, IIA6R6, and IA8R8 potentials, respectively. In terms of polydispersity though, a notable increase relative to the IBI potential is evident. This discrepancy may be ascribed to the ability of the IBI optimization to tune the positions in $r$ of the primary attractive well and repulsive barrier independently of their ranges, allowing for greater precision in the repulsive shell surrounding the cluster. Nonetheless, the above results demonstrate that only a single attractive well followed by repulsive barrier are strictly necessary for assembly of clusters with a reasonably well-controlled size for potentials of this type. 

\begin{figure}
  \includegraphics{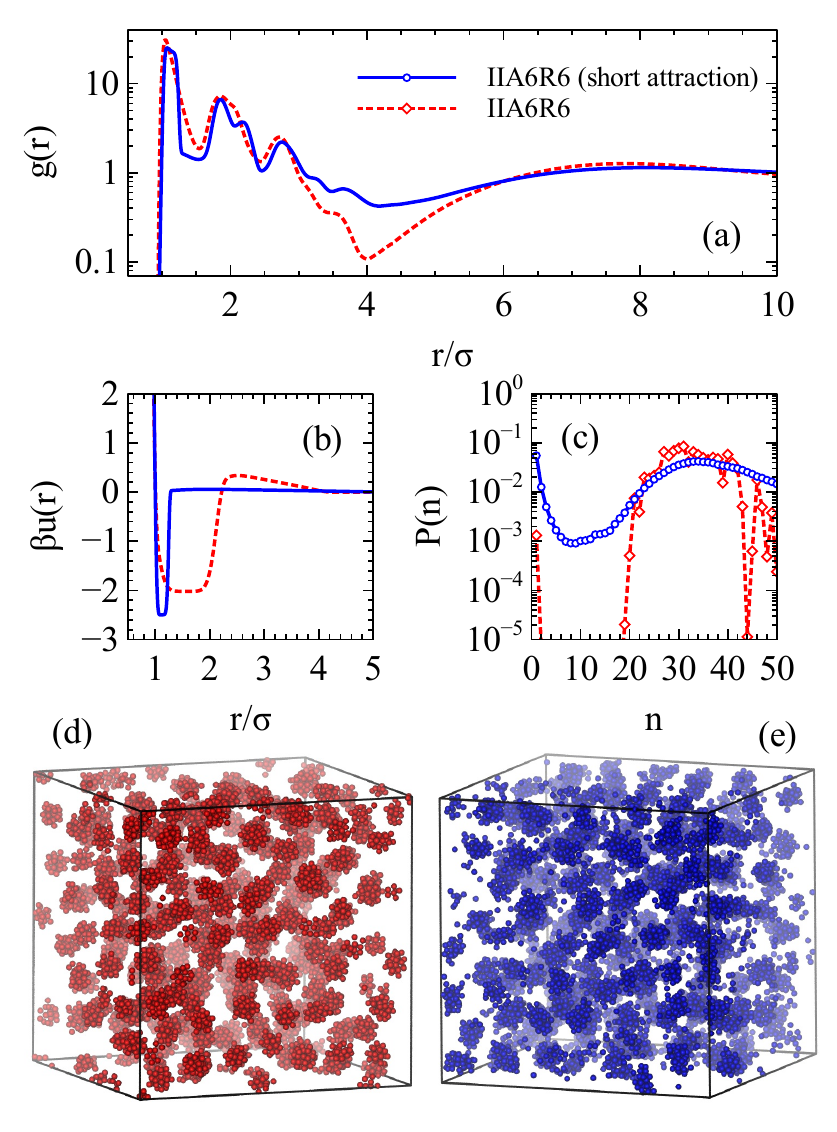}
  \caption{For an equilibrium clustered fluid target (a) the radial distribution functions, (b) dimensionless pair interactions, $\beta u(r)$, (c) their associated CSDs, and representative particle configurations for the IIA6R6 functional form obtained via RE in the absence (d) or presence (e) of a constraint on the breath of the attractive well (fixed at $0.25\sigma$).}
  \label{fgr:FigureSALRComparison}
\end{figure}

While this new interaction class seems to provide generally enhanced clustering relative to many previously reported SALR interactions, it is possible that a more traditional SALR interaction form could also lead to more idealized clusters if purposefully engineered via ID to do so. To address this we have optimized the IIA6R6 potential with the attraction range constrained to $0.25\sigma$. As a consequence of this additional constraint, the intra-cluster droplet structure deviates slightly and the interface region between clusters is less strongly depleted of particles (see the radial distribution functions in Fig.~\ref{fgr:FigureSALRComparison}a). Furthermore, the repulsion optimizes to a more canonical SALR form with a weak, long ranged repulsive ramp, as shown in Fig.~\ref{fgr:FigureSALRComparison}b. As evident from the CSDs in Fig.~\ref{fgr:FigureSALRComparison}c, clustering is much weaker in the constrained IIA6R6 system when compared to the unconstrained analog. Specifically, enhanced monomer is found in the former in tandem with a weaker monomer-cluster bimodality. Despite the overall decreased performance with constraining the width of the attractive well, the ID methodology is still able to closely target the desired cluster size, achieving a mean aggregate size of $\langle n \rangle=34.4$, though polydispersity is visibly enhanced. The lower quality of clustering is also evident visually by comparing snapshots of the unconstrained and constrained IIA6R6 systems (Fig.~\ref{fgr:FigureSALRComparison}d and e, respectively). 

On the whole, there are several advantages in extending inverse calculations from an unrestricted potential (as in IBI) to a potential with a prescribed functional form. Though the IBI calculations are valuable to discern whether it is at all possible to assemble a desired structural motif with a pair interaction, it is not realistic to expect that every detail present in the unconstrained interaction is realizable in an experimental system. By contrast, the functional forms encoded here possess only two characteristic length scales that are coupled in $r$. It is conceivable to combine one source of interparticle attraction (e.g., depletion,~\cite{CollInt1} hydrophobic effect~\cite{hydrophobic_effect}) and one source of repulsion (e.g., electrostatics,~\cite{CollInt1,HansenMcDonald} sterics~\cite{CollInt1}) in a colloid- or nanoparticle-based to arrive at qualitatively similar interparticle potentials. Ultimately, the discovery of multiple, relatively simple pair interactions that form clusters sets the stage for the use of physics-based models in RE calculations, where physically motivated constraints can be applied to parameters that bear direct relation to experimental conditions.

\section{Pores}
\label{sec:pores}

\subsection{Target Simulation}
In Ref.~\citenum{pores} and~\citenum{pores2}, target ensembles of particles with morphologies consistent with a porous mesophase were fabricated via an MD simulation of a binary hard sphere-like mixture with a large size asymmetry (here, a 4:1 diameter ratio). Large particles--which provided spherical ``pore'' zones of excluded volume--were fixed to FCC lattice positions in a sea of mobile small particles that constitutes the surrounding matrix. Within this framework, the target ensemble is regarded as the set of small particle configurations generated via the molecular simulation in the presence of the immobile exclusion spheres. Based on our prior work on clusters that demonstrated the importance of constructing a target simulation that can be reproduced by pair interactions,~\cite{ID_clusters} we placed the large spheres on a lattice with a nearest neighbor distance such that the thickness of the matrix surrounding the pore is commensurate with the pore diameter. By choosing a pore nearest neighbor distance to pore diameter ratio of 1.85, most particles will be positioned such that a repulsive barrier falling near to the pore diameter will function to stabilize multiple pores.~\cite{pores} This is, again, a consequence of the isotropic nature of the pair interactions, such that the repulsive barrier helps to forms a spherical corona around the particle center. 

\subsection{Prior work}

Unlike clustered fluid phases--which have been extensively studied, especially within the context of SALR pair potentials--self assembly of the conjugate ``inverse'' cluster phase (also known as the bubble, void, or pore phase) has largely remained elusive, especially in 3D. Intriguing theoretical studies suggested the possibility of pore formation with a simple pair potential comprising competitive attractive and repulsive components.~\cite{postulated_phases_sear,mean_field_assembly_0,mean_field_assembly_1,mean_field_assembly_2} Further support for this notion has come in the form of 2D lattice models~\cite{lattice_bubbles_1, lattice_bubbles_2} and continuum field theories~\cite{2D_microphase}, and experimental, quasi-2D interfacial colloid assembly.~\cite{jaime_expt,ghezzi_expt} Prior to the application of ID, however, the closest simulated 3D pore assembly resulted in a single large void within a simulation cell, as specific control over the pore size was not yet possible.~\cite{simulated_phases} Recently, leveraging ID, we discovered interactions that generate an assembly of spherical pores with a prescribed size.~\cite{pores} As shown below, the unconstrained IBI optimization yielded an interaction with competing attractive and repulsive components, but also with deviations from the canonical SALR model in a similar fashion to the previously discussed cluster interactions. The previous study also demonstrated the relative insensitivity of the porous assembly to the secondary features in the pair potential: the same constrained functional forms employed in Sect.~\ref{subsec:cluster_results} were found to assemble porous structures upon optimization with RE.

Application of the proposed ID methodology not only enabled the discovery of a 3D assembly of pores, but also yielded rich phase behavior akin to that found in block copolymers: by reducing the packing fraction, a phase progression from pores$\rightarrow$void columns$\rightarrow$bi-continuous$\rightarrow$lamellar$\rightarrow$particle columns$\rightarrow$clusters was discovered,~\cite{pores} validating an array of early theoretical predictions and suggestive simulations using competing interactions. 

\subsection{Results}

\begin{figure}
  \includegraphics{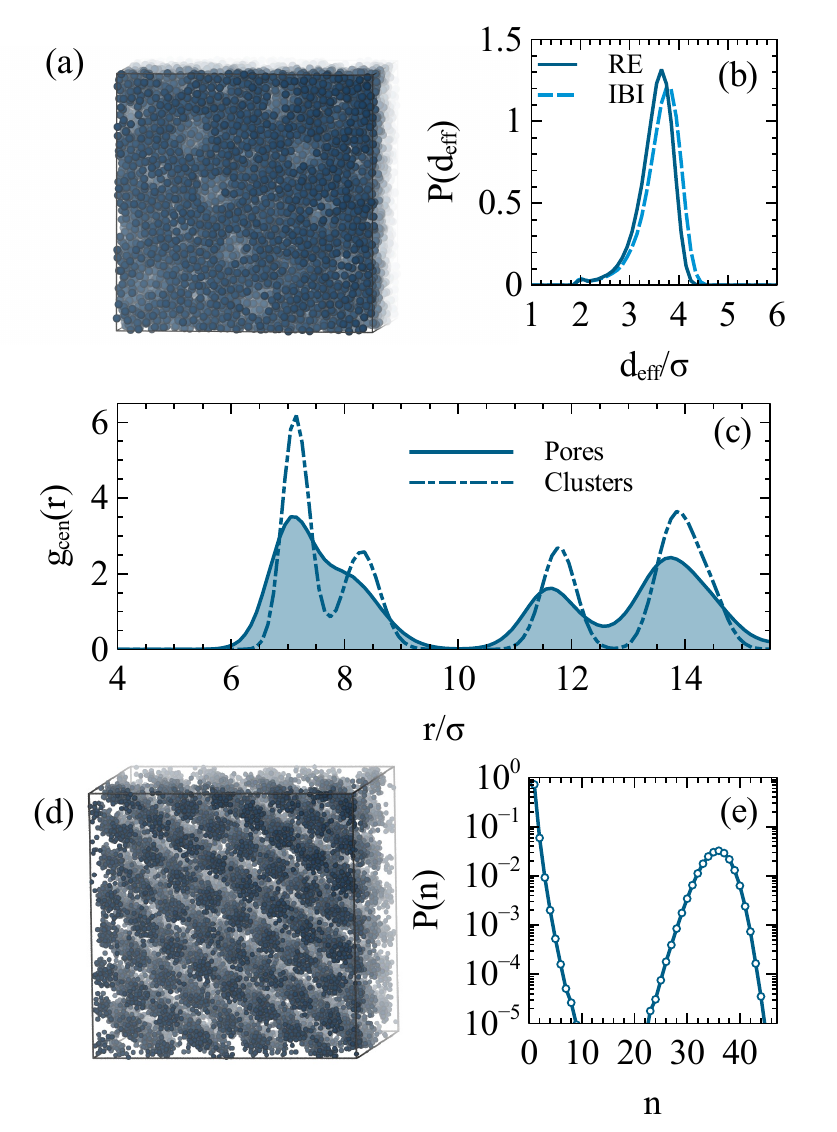}
  \caption{(a) Representative configuration of a porous assembly generated by the purely repulsive pair interaction obtained via RE, and (b) comparison of the pore size distributions (PSDs) resulting from RE and IBI. (c) The pore-pore and cluster-cluster $g(r)$ with the RE repulsive pair interaction at packing fractions of 0.31 and 0.07, respectively. (d) Representative configuration of the ordered clustered phase from (c), and (e) the CSD at this state point.}
  \label{fgr:RepPoreStats}
\end{figure}

Prior work has shown that competing attractions and repulsions are \emph{not} strictly required in a pair potential to allow particles to form a microphase segregated structures: some purely repulsive interactions have been predicted to stabilize clustered phases, for instance.~\cite{likos_repcluster,camp_repclusters,frenkel_repmicrophases} Via application of RE minimization, we discover here a purely repulsive pair potential that assembles particles into a porous mesophase, as can be seen in Fig.~\ref{fgr:RepPoreStats}a.
In Fig.~\ref{fgr:RepPoreStats}a, lighter spherical-like regions from which particles are absent are interspersed among a fluid of dark blue particles.
(The technical implementation of the pair potential constraint applied in the RE optimization is discussed in the Appendix.) Despite the absence of an attractive component, reasonable control over the pore size is achieved; the pores are only slightly smaller than the pores targeted or generated via IBI. Moreover, polydispersity is not significantly altered with respect to the unconstrained (IBI) structures (Fig.~\ref{fgr:RepPoreStats}b). The pores are arranged on a BCC lattice; the pore-pore $g(r)$ is shown in Fig.~\ref{fgr:RepPoreStats}c. 

In addition to achieving pore assembly, the purely repulsive pair interaction also exhibits a qualitatively similar block copolymer-like phase progression discovered in the unconstrained system, with all of the same microphase-segregated states (clusters, columns, lamellae) present at some $\eta$. An image of the clustered phase is shown in Fig.~\ref{fgr:RepPoreStats}d; this structural motif emerges near $\eta=0.07$--identified as value of $\eta$ for which the CSD indicates the strongest preference for clustering (shown in Fig.~\ref{fgr:RepPoreStats}e). Compared to the clustered fluids explored in Sect.~\ref{sec:clusters}, there is a greater degree of monomer and other small aggregates present, which reduces the magnitude of the peak associated with the clusters in the CSD. However, the breadth of the main peak in the CSD indicates that the monodispersity of the clusters is comparable to that seen in Sect.~\ref{sec:clusters} for the unconstrained (IBI) potential.

Visual inspection of the clustered configuration in Fig.~\ref{fgr:RepPoreStats}d reveals that, unlike in Sect.~\ref{sec:clusters}, the clusters are arranged on a lattice. Interestingly, comparison of the peak positions in the cluster-cluster and pore-pore radial distribution functions in Fig.~\ref{fgr:RepPoreStats}c indicates that the microphase-separated objects (either clusters or pores) are positioned in essentially the same crystalline structure: a BCC lattice with approximately equal nearest neighbor distance. It is interesting that the general pore-cluster symmetry extends to such quantitative details as well.

\begin{figure}
  \includegraphics{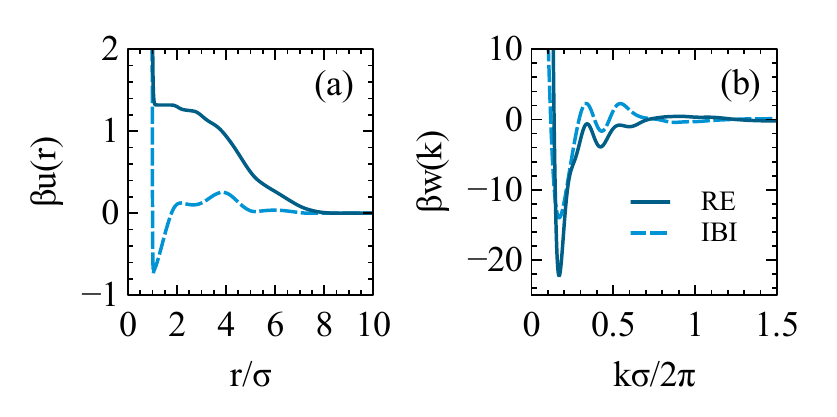}
  \caption{(a) Pair interactions, $\beta u(r)$, and (b) their Fourier space representations, $\beta w(k)$, optimized with ID for a porous target for the unconstrained (IBI) and purely repulsive interaction (RE) cases. See the main text for the definition of $w(r)$.}
  \label{fgr:RepPorePotential}
\end{figure}

The ability of a purely repulsive interaction to elicit nearly identical assembly to one with competing attractive and repulsive contributions may seem surprising (see Fig.~\ref{fgr:RepPorePotential}a); however, examining the potentials in Fourier space reveals hidden similarities. Minima in the Fourier transform of a pair interaction, $u(k)$, indicate energetically preferred length scales for structural correlations, as a minimum in $u(k)$ promotes a corresponding maximum in the structure factor, $S(k)$, at the same wave-vector $k$.~\cite{fourier_analysis} In this case, the analysis requires extracting the non-core, external contribution to the potential, defined here as $w(r)\equiv0$ for $r\leq\sigma$ and $w(r)\equiv u(r)$ for $r>\sigma$. The core contribution is omitted because (1) the core amplitude is very large and will otherwise dominate the Fourier transform, and (2) we are interested in how the potential will modify structure in excess of the geometric constraints imposed via a steeply repulsive core. The Fourier transform of $\beta w(r)$, $\beta w(k)$, is shown in Fig.~\ref{fgr:RepPorePotential}b, where it is clear that the low-$k$ behavior of both potentials is rather similar. Both have a strong energetic preference towards intermediate range (IR) wave-vectors, the two lowest of which are at virtually identical $k$. While the presence of energetically preferred IR wave-vectors in $\beta w(k)$ is not sufficient to guarantee pore formation, if pores are assembled, they should have similar packing correlations and characteristic length scales--as we have found here to be true. 

The capability of a purely repulsive interaction to elicit almost identical behavior to an unconstrained counterpart is a testament to the interchangeability of attractions with regions of low force in the interaction and augmented pressure. As can be seen in Fig.~\ref{fgr:RepPorePotential}a, the repulsive constraint yields plateaus in the pair potential over distances whereby the unconstrained analog has attractions. Such plateaus are even more obvious in repulsively assembled crystals of the next Section. Coupled with external applied pressure to maintain an appropriate density, these plateaus yield preferential particle-particle packing length scales, akin to attractions, as the rest of the potential acts to force particles apart. 

\section{Crystals}
\label{sec:crystals}

\subsection{Target Simulation}

As in Ref.~\citenum{BL_RE}, target crystal ensembles were created by tethering \emph{non-interacting} particles to the desired lattice positions via a harmonic constraint in an MD simulation. Relatively strong spring constants were employed (i.e., the ``temperature'' of the simulation was cold), yielding highly localized crystalline configurations and sharp coordination shell structure in $g(r)$. Strong localization (1) helps ensure that a stable solution to the optimization exists and (2) mitigates, or removes entirely, any competition from a non-crystalline (fluid) solution to the maximization. However, some fluctuations about the lattice positions are necessary such that the $g(r)$ is integrable in the RE update scheme.

\subsection{Prior Work}
\label{subsec:xtal_prior}

Research on self-assembly of crystalline lattices from pair potentials has spurred the development of many ID strategies, most of which employ a ground state (zero-temperature) calculation wherein the target lattice is stabilized relative to an explicit pool of competitor crystals.~\cite{ST_inv_des_review,AJ_inv_des_review,ID_crystals_TS,ID_crystals} Neither of these attributes is optimal, as (1) ground state stability does not guarantee any appreciable stability upon heating, and (2) a complete pool of competitor crystals can only be asymptotically approached. As a consequence, post optimization simulation of the designed interaction often yields an undesired structural motif that was not included in the competitor pool--necessitating another optimization with an augmented pool. Despite these limitations, such approaches have had notable successes leading to the discovery of purely repulsive pair interactions capable of stabilizing a diverse array of complex, low coordinated crystals in 2D and 3D.


\begin{figure}
  \includegraphics{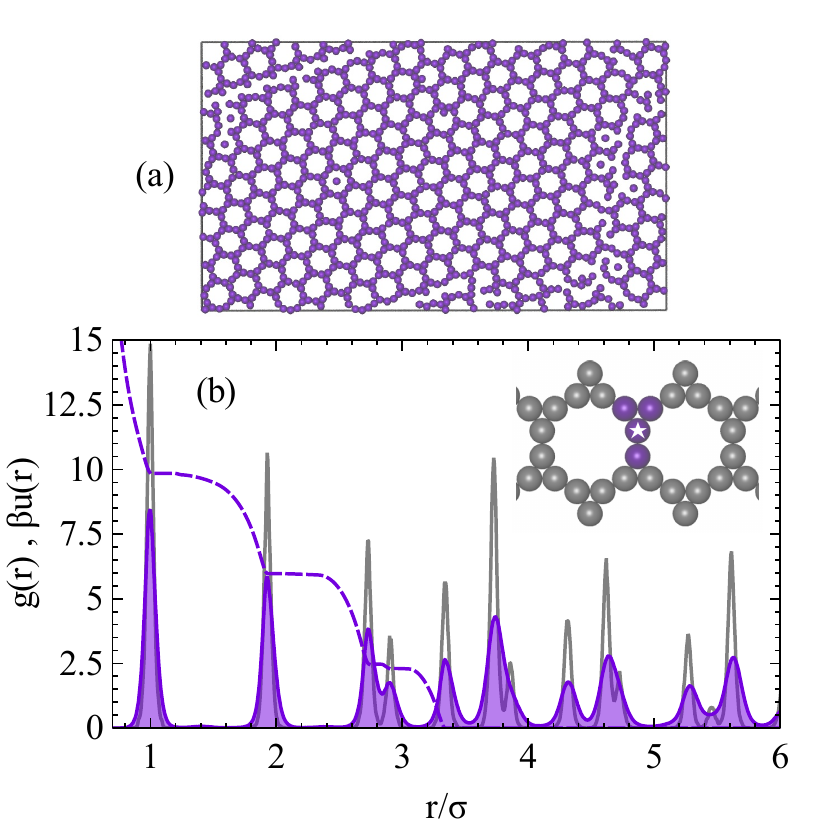}
  \caption{For the TH lattice, (a) a representative configuration at the end of the ID optimization, and (b) the ID optimized pair interaction, $\beta u(r)$ (dashed), and comparison of $g(r)$ given by $\beta u(r)$ (filled) and $g_{\text{tgt}}(r)$ (no fill). The inset depicts the ideal TH lattice positions, where the first coordination shell of a selected particle (star) is highlighted in purple.}
  \label{fgr:FigureTH}
\end{figure}

The present methodology utilizes the finite-temperature MD simulations performed at each optimization step to provide an implicit pool of competitors via the self-assembled structures arising during the MD simulations as well as the fluid state that exists prior to crystallization. Once assembled however, lattices can possess significant kinetic stability even if they are not thermodynamically stable. Therefore, after self-assembly occurs within the optimization, the resulting potentials from future iterative steps are no longer guaranteed to self-assemble into the desired structure from a fluid. To remedy this, each new optimization step included a heating cycle to melt the crystal, followed by a slow cooling back to the working optimization temperature. Recently, this methodology was employed to discover a variety of purely repulsive pair interactions that each self-assemble particles into a two-dimensional lattice in finite-temperature MD simulations.~\cite{BL_RE} Honeycomb, kagome, square, rectangular, truncated square, and truncated hexagonal crystals were all successfully assembled via ID potentials; the latter two structures had not been realized previously with any pair interaction.

\begin{figure}
  \includegraphics{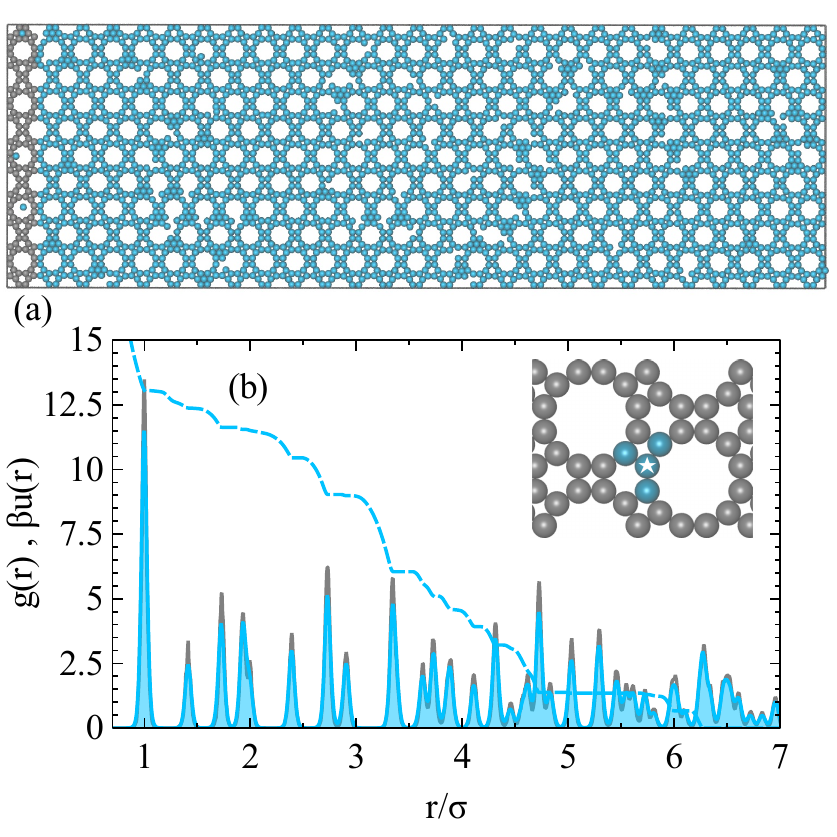}
  \caption{For the TTH lattice, (a) a representative configuration at the end of the ID optimization where the seeded particles are shown in gray, and (b) the ID optimized pair interaction, $\beta u(r)$ (dashed), and comparison of $g(r)$ given by $\beta u(r)$ (filled) and $g_{\text{tgt}}(r)$ (no fill). The inset depicts the ideal TTH lattice positions, where the first coordination shell of a selected particle (star) is highlighted in blue.}
  \label{fgr:FigureTTH}
\end{figure}

\subsection{Results}
Issues of practical realizability aside, we wish to explore the limits of crystal self assembly within the proposed methodology. For context, we first recall the truncated hexagonal (TH) lattice (shown in Fig.~\ref{fgr:FigureTH}a) assembled in Ref.~\citenum{BL_RE}. While this lattice is composed of identical sites, it is challenging to self assemble with an isotropic pair interaction given the high distributional inhomogeneity of particles about one another--yielding only two-fold reflection symmetry, as well as a high degree of angular asymmetry, in the first coordination shell (see inset of Fig.~\ref{fgr:FigureTH}b). Moreover, there are a variety of competitor phases that are relevant, owing to the complexity and openness of the lattice. 
In particular, an alternative ID approach based on ground state computations required the presence of 41 distinct structural motifs in the competitor pool to stabilize the TH lattice.~\cite{WP_TH} Despite these difficulties inherent to the TH lattice, the ID scheme easily discovered an appropriate repulsive pair interaction for assembly of the TH lattice; an interaction possessing three broad features was required, spanning five coordination shells (see Fig.~\ref{fgr:FigureTH}b). 

A more complex target lattice is the truncated trihexagonal (TTH) lattice (shown in Fig.~\ref{fgr:FigureTTH}a). The TTH lattice possesses no symmetry whatsoever in the first coordination shell (see inset of Fig.~\ref{fgr:FigureTTH}b). 
To our knowledge, no lattice possessing $\emph{C}_{1}$ symmetry with respect to its first coordination has ever been assembled via an isotropic pair interaction. Direct application of the approach outlined in Sect.~\ref{subsec:xtal_prior} yielded assemblies that, while locally bearing resemblance to structural elements present in the TTH lattice, were malformed and devoid of any TTH unit cells.

Suspecting that slow nucleation kinetics were to blame for the failure to assemble the TTH lattice, we integrated a crystalline seed into the optimization protocol to accelerate self-assembly.~\cite{torquato_seed,nucleation_id} The seeded particles are shown in gray in Fig.~\ref{fgr:FigureTTH}a. To ensure that the simulated $g(r)$ is smooth and integrable, the seeded particles are not frozen but rather are tethered to lattice positions via a \emph{weak} harmonic constraint. Because the method ultimately requires only the simulated $g(r)$ to update the interparticle interactions, it has the advantage that the optimization scheme remains valid under a variety of simulation protocols (e.g., constraints, fields), so long as the resulting $g(r)$ is integrable over the range of potential. 
With the crystalline seed, the ID methodology discovered a pair interaction that assembles into TTH. Ultimately, self assembly of TTH required that a minimum of $~19$ coordination shells be encompassed by the pair interaction (Fig.~\ref{fgr:FigureTTH}b). While such as interaction is highly unlikely to be physically realizable, the complexity of potential required for self-assembly of TTH is a testament to both the difficulty of the target and the robustness of the ID scheme. 
%

The ability to incorporate a seed in the ID scheme is a valuable asset both in terms of accelerating the \emph{in silico} assembly of lattices with slow nucleation kinetics as well as in relation to experimental work that may intentionally employ crystalline seeds. However, implications on the thermodynamic stability for the desired lattice in the absence of the seed are unclear. To address this, we generated a small ensemble of unseeded, self-assembled structures by slowly cooling five simulations from a disordered fluid to $T=0$ with particles interacting via the potential from Fig.~\ref{fgr:FigureTTH}. The structure shown in Fig.~\ref{fgr:FigureTTH} was similarly cooled in the absence of the harmonic constraints on the seed particles. In the absence of a seed during self-assembly, highly malformed structures developed, possessing an overabundance of hexagonal motifs and a significant dearth of dodecagons. The seeded structure, on the other hand, remains as a TTH lattice, even after the harmonic constraints on the seed particles were removed. The malformed assemblies are uniformly higher in energy than the TTH lattice, strongly suggesting that the ground state for this potential, with or without the seed, is TTH.

Despite large structural differences between the seeded and unseeded assemblies, the difference in energy at $T=0$ is small, with the malformed assemblies only $~0.7\%$ higher in energy. With such subtle differences between the correct assembly and the multitude of incorrect alternatives, the role of the seed is clear. In essence, it provides a slight bias towards the desired structure in the vast landscape of malformed configurations that are similar in energy. Besides the computational benefits of seeded ID, this technique is also of direct relevance to substrate and template-assisted growth whereby the ``seed'' takes on a literal role.

\begin{figure}
  \includegraphics{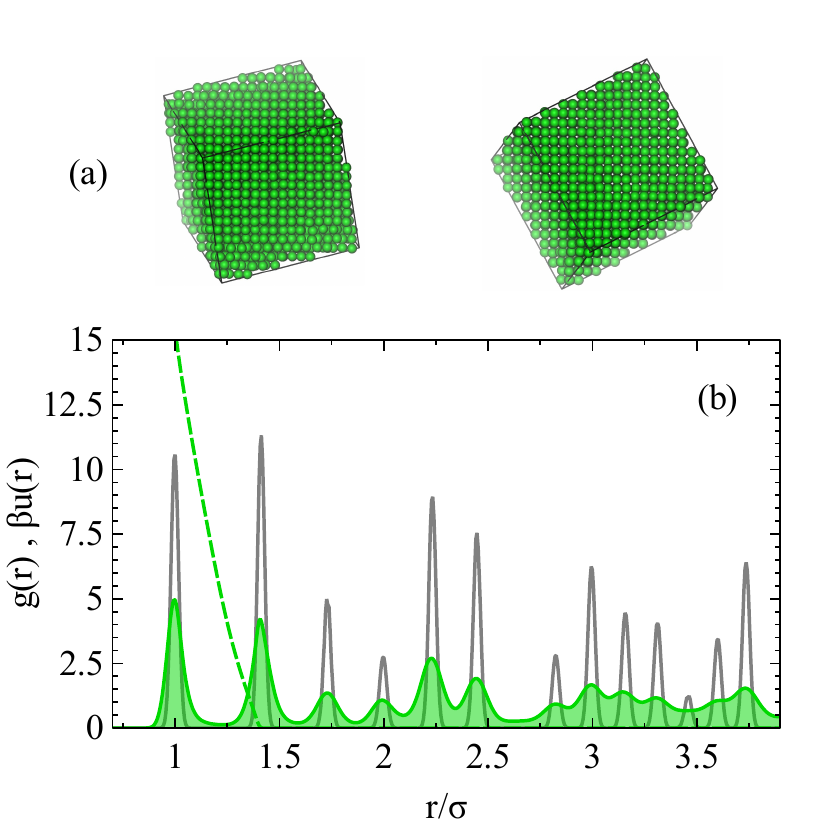}
  \caption{For the SC lattice, (a) two views of a representative configuration at the end of the ID optimization, and (b) the ID optimized pair interaction, $\beta u(r)$ (dashed), and comparison of $g(r)$ given by $\beta u(r)$ (filled) and $g_{\text{tgt}}(r)$ (no fill).}
  \label{fgr:SC}
\end{figure}

\begin{figure}
  \includegraphics{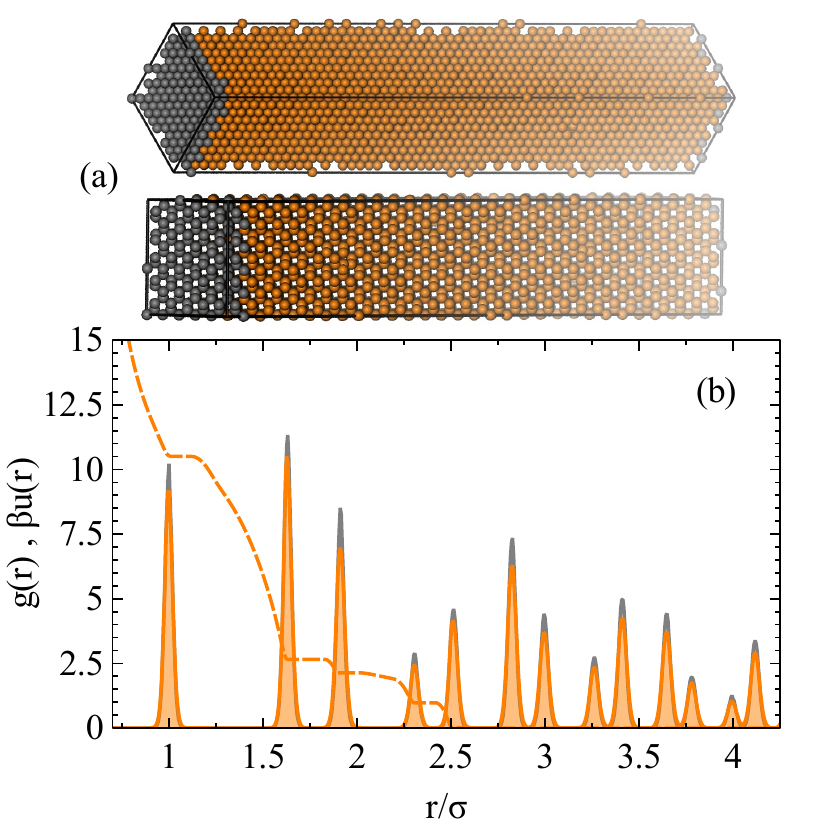}
  \caption{For the diamond lattice, (a) two views of a representative configuration at the end of the ID optimization where the seeded particles are shown in gray, and (b) the ID optimized pair interaction, $\beta u(r)$ (dashed), and comparison of $g(r)$ given by $\beta u(r)$ (filled) and $g_{\text{tgt}}(r)$ (no fill).}
  \label{fgr:Diamond}
\end{figure}

Because our ID scheme for crystals requires repeated on-the-fly self-assembly of the desired microstructure, slow nucleation kinetics are of particular concern for application to 3D lattices. As dimensionality increases, in general nucleation kinetics slow and more particles are needed to mitigate potential finite size effects.  
Despite these issues, we discovered via ID a pair interaction that quickly and accurately self-assembles into a simple cubic crystal in the absence of a seed. As shown in Fig.~\ref{fgr:SC}, assembly is almost perfect--only a small defects arises due to incorrect alignment with the simulation box. As with the square lattice (the 2D analogue to simple cubic), the simple cubic lattice is realized via a relatively simple potential comprised of a repulsive core and single shoulder spanning the first and second coordination shells. Another common target in 3D crystal ID is the diamond lattice which, not surprisingly, we find more difficult to assemble given its relatively open and complex structure. Unseeded ID attempts led to malformed assemblies. Leveraging seeded growth, however, yields a well-assembled diamond lattice with the potential shown in Fig.~\ref{fgr:Diamond}. However, five coordination shells of structure were required to thwart the formation of non-diamond structures bridging the propagating crystalline fronts. 
Similarly to TTH, the optimized diamond potential did not form the correct lattice in the absence of a seed, in this case forming amorphous glassy states. Using a similar protocol to that employed for TTH above (see Appendix), we confirmed that the glassy states are $\approx 10\%$ higher in energy than the seeded crystalline morphology, suggestive that slow kinetics for assembly thwart crystallization in the absence of a seed. 



\section{Conclusions}
\label{sec:conclusions}

In this article we outlined a general, statistical mechanics based approach for the ID of interactions resulting in self assembly and applied the methodology to discover isotropic, pair interactions that formed clusters, pores, and crystalline lattices at equilibrium. These applications emphasize a potential advantage of the method, wherein the discovery of these interactions is recast from a forward search through parameter space to the design of an appropriate target simulation to represent a desired structural motif. That is, in forward design studies, parameters must be tuned (often empirically) with the hope of eliciting a desired behavior; however, the number of parameters can be quite large and each can have a complex and non-intuitive effect on the structural properties. By leveraging ID, these challenges are forgone in favor of the challenges inherent to designing a suitable target ensemble. Often, these latter challenges are more amenable to physical intuition as opposed to brute force. For example, in the design of clusters and pores, judicious choice of a single length scale in the problem (the thickness of the repulsive shell around clusters the pore-pore spacing, respectively) in the target simulations had profound effects on the realizability of each with a pair potential.

Having demonstrated the application of the proposed ID methodology in a variety of contexts, the stage is set to explore various extensions. For instance, of great importance for experimental practicality is the discovery of interactions that exhibit the desired structural motifs over larger ranges in particle concentration. Two directions seem profitable towards this end. The first is an approach borrowed from the machine learning community: flexible hyper-parameters (here, tunable parameters potential parameters) could be regularized to maximize the model generality to conditions for which it was not trained. For example, penalizing large Akima spline step parameters (effectively penalizing large forces) would inevitably soften the observed shoulders for the lattice interactions, likely improving the density stability range. This would almost certainly come at the cost of diminished temperature stability,~\cite{crystal_temp_stability} so a compromise would have to be identified. The second, more direct option is to leverage multi-state ideas from the biomolecular coarse-graining community. In this scenario the probabilistic formalism must be extended to maximize the likelihood of realizing target configurations at an array of state points. 

Beyond improving the flexibility of designed interactions, it is desirable to extend the approach to leverage anisotropic interactions and to directly target material properties as opposed to self-assembly. In regards to the former, patchy models would serve as a tractable starting point; possible variables include patch location, net attraction or repulsion between patch pairs, and the number of patches. 
Though widely studied via forward searches,~\cite{patchy_review1,patchy_review2,kagome,capsid,BCPs} self-assembly of anisotropic particles has not been heavily investigated via inverse methods, making the area particularly fertile ground for both validation of new ID methods as well as discovery of novel anisotropic interactions.~\cite{nucleation_id,patchy_screening,template_directed} 
For the latter an endless array of properties can be explicitly optimized, including the elastic modulus, viscosity, terminal jamming density upon compaction, single electron hopping conductivity, percolation threshold and even fractal dimension of a percolated particle network structure. Such developments would nicely complement the self-assembly oriented approach outlined in this article.

Finally, we emphasize that the array of targets employed in this work is far from exhaustive; the only limit pertains to the creativity needed to construct a target ensemble and its realizability with a given potential interaction form. Only out of convenience did we adopt statistical physics based models for $P_{\text{tgt}}(\textbf{R})$ (i.e., states are sampled according to a Boltzmann weight with some fictive temperature and many-body potential). 
This need not be the case. Any means of generating an ensemble of configurations is valid as long as it has convergent large sample statistics (even if $P_{\text{tgt}}(\textbf{R})$ is not explicitly known for the protocol). For example, random sequential addition (RSA) is a highly studied, non-equilibrium approach to packing spheres. It proceeds via sequential, individual random sphere additions to a volume with rejection if an overlap is generated. This yields a well-defined terminal packing density at which no more spheres can be added. Using ID one could discover interactions that, at a fixed density and thermodynamic equilibrium, sample similar configurations to those in RSA.

\section*{Acknowledgments}
This work was partially supported by the National Science Foundation (1247945) and the Welch Foundation (F-1696 and F-1848). We acknowledge the Texas Advanced Computing Center (TACC) at The University of Texas at Austin for providing HPC resources.

\setcounter{figure}{0}
\setcounter{equation}{0}
\renewcommand\thefigure{A\arabic{figure}}
\renewcommand{\thesection}{\thepart .\arabic{section}}
\renewcommand\theequation{A\arabic{equation}}
\renewcommand{\thesubsection}{\arabic{subsection}}
\renewcommand{\thesubsubsection}{\alph{subsubsection}}

\section*{Appendix}

\subsection{Constrained potential forms}

In Section~\ref{sec:clusters}, we leveraged RE minimization and constrained pair interactions to design potentials that assemble a fluid phase of equilibrium clusters, the definitions of which are 
\begin{equation} \label{eqn:re_potential}
\begin{split}
&u_{\text{RE}}^{(i)}(r) \equiv -\epsilon_{1} \textup{exp}\bigg[-\bigg(\dfrac{r-d-\alpha_{1}/2}{\alpha_{1}/2}\bigg)^{n_{1}}\bigg] \\ &+  \epsilon_{2}q_{i}(\alpha_{1},\alpha_{2}) \textup{exp}\bigg[-\bigg(\dfrac{r-d-\alpha_{2}/2-\alpha_{1}}{\alpha_{2}/2}\bigg)^{n_{2}}\bigg]
\end{split}
\end{equation}
where 
\begin{equation} \label{eqn:re_potential_2}
q_{i}(\alpha_{1},\alpha_{2})\equiv \delta_{i,\text{I}} + \delta_{i,\text{II}}\bigg[ \dfrac{-r + d+\alpha_{1}+\alpha_{2}}{\alpha_{2}} \bigg]
\end{equation}
\(\delta_{i,j}\) is the Kronecker delta, [\(\epsilon_{1},\epsilon_{2}\)] are the [attractive, repulsive] strengths, and [\(\alpha_{1},\alpha_{2}\)] are the corresponding ranges. The extra multiplicative factor in the type II potential approximately converts the symmetric repulsive hump into a linear ramp. The potentials also contain two non-optimized exponents, [\(n_{1},n_{2}\)], the magnitude of each controls how square-like the attraction and repulsion are respectively. For convenience, we adopt the compact notation, \(i\text{A}n_{1}\text{R}n_{2}\), to convey the interaction type and the exponents employed. 


The assembly of purely repulsive porous and crystalline phases in Sect. ~\ref{sec:pores} and ~\ref{sec:crystals}, respectively, employed a constrained Akima spline with $n$ knots at locations $\{r_{1}, r_{2}, ..., r_{n}\}$, where $r_{n}$ is the cut-off for the potential, and a corresponding set of potential amplitudes $\{u_{1}, u_{2}, ..., u_{n} \}$, which the Akima spline interpolates through. At each knot we define a tunable parameter as $\theta_{i}=u_{i}-u_{i+1}$ (the difference in amplitude at the current knot relative to the next knot), the entire set of which defines our parameter optimization vector $\boldsymbol{\theta}=[\theta_{1}, \theta_{2}, ..., \theta_{n-1}, u_{n}]$. Adopting this parametrization allows us to easily enforce monotonicity (by demanding all $\theta_{i}\geq0$) and a smooth force cutoff (via $\theta_{n-1}=0$ and $u_{n}=0$). 

As our starting guess for the Akima spline optimization we utilize a power law form (weighted by a step-like smoothing function to minimize the force near and at that cutoff) to assign the initial $\boldsymbol{\theta}$ based on the set of knot locations corresponding amplitudes extracted via
\begin{equation}
\beta u_{0}(r) \equiv \dfrac{1}{2} A\bigg(\dfrac{\sigma}{r}\bigg)^{a} \bigg(1-\tanh \bigg[k\bigg(\dfrac{r}{\sigma} - \dfrac{r_{s}}{\sigma}\bigg)\bigg]\bigg)
\end{equation}
where $A$ is the dimensionless amplitude, $k$ controls how rapidly the potential dies off, $a$ is the power, $r_{\text{s}}$ is the center of the smoothing function, and $\sigma$ is the nearest neighbor crystal distance. As in a previous publication, we set $A=1.8$, $k=8.9$,  $a=5$ and $r_{\text{s}}$ is set close to, but before, the potential cutoff. Empirically, we find that the optimization is not sensitive to the precise initial guess. 

We employed the Akima spline $u(r)$ for the ID of both pore and crystal phases using respective knot spacings of $0.022$ and $0.017$. For pores, a cutoff of $8.0\sigma$ was used, whereas for crystals the cutoff was systematically chosen to include as few coordination shells as possible while still allowing for the correct lattice to form. The specific cutoffs used for truncated hexagonal, truncated trihexagonal, simple cubic, and diamond correspond to $3.49\sigma$, $6.28\sigma$, $1.56\sigma$, and $2.59\sigma$ respectively. For both pores and crystals, the cutoff knot locations are followed by two additional unoptimized knots with parameters set to zero to ensure a smooth vanishing of the force.

\subsection{Additional simulation details}

\subsubsection{Clusters}

Here we provide additional cluster simulation details regarding the assembly simulations only; specific details pertaining to the target structure can be found in Ref.~\citenum{ID_clusters}.

Optimizing clusters required a significant number of iteration steps; however, each iteration is relatively inexpensive due to the low volume fraction and thus minimized number of interacting neighbors. In all cases $8000$ particles were simulated at each iteration and a total of 500 to 1000 iterations were necessary to reach satisfactorily small updates to the pair potential and corresponding structure. Each iteration employed a simulation of order $1\times 10^{6}$ MD steps. The final $g(r)$ and CSD analysis used another $2\times 10^{6}$ equilibration steps and $2\times 10^{7}$ steps to accrue enough statistics to yield a broad temporal sample of the slow cluster exchange--of particular relevance to the CSD calculation.

\subsubsection{Pores}

For the porous structure, converged results were achieved via 40 optimization steps, each comprised of $8\times 10^5$ MD steps. With the optimized potentials in hand, production runs of $11\times 10^{6}$ steps were carried out, where configurations were sampled from the last $6\times 10^{6}$ steps to generate statistics for the $g(r)$ and size distributions associated with the pores and clusters. Relative entropy calculations were performed with 5475 particles, and characterization was performed with 18478 particles--chosen so that at the optimization density, the box size was 1.5 times larger than in the optimization to check for finite size effects. More details on the target simulation for the pores can be found in Refs.~\citenum{pores} and~\citenum{pores2}. 

\subsubsection{Crystals}

Here we provide additional crystal simulation details pertaining to both the target and assembly simulations. The 2D truncated hexagonal (TH) and truncated trihexagonal (TTH) crystal simulations contained $1200$ and $3240$ particles respectively, with corresponding box dimensions of $37.27\sigma\times64.55\sigma$ and $40.92\sigma\times127.59\sigma$. 3D simulations of simple cubic (SC) and diamond (DIA) crystals employed $4096$ and $6000$ particles respectively, with simulation box dimensions of $15.98\sigma \times 15.98\sigma \times 15.98\sigma$ and $11.53\sigma \times 11.53\sigma \times 69.19\sigma$. In both the target and seeded ID simulations, harmonic constraints were employed to tether particles to specified crystalline coordinate positions. Specifically, target simulations employed relatively strong spring constants of 2237, 2796, 6408, and 4793 $k_{B}T/\sigma^{2}$ for the TH, TTH, SC, and DIA crystals respectively, yielding a nearest-neighbor peak in $g(r)$ with a magnitude of order 10. For the ID optimization of the TTH and DIA crystals, we used seeds comprised of 120 and 250 particles respectively. They were designed to perfectly fill space in the short dimension(s) and provide one unit cell of structure in the long dimension. Like the target simulations, seeds utilized a harmonic constraint to maintain the crystal structure; however, the seed tethers were comparatively weak, possessing spring constants of $~40k_{B}T/\sigma^{2}$ and $~801k_{B}T/\sigma^{2}$ for TTH and DIA respectively. The weaker restoring forces help to minimize the contribution to localization upon crystallization.

The computational demand of the optimization protocol can vary significantly from one crystal to another. For less challenging crystals, like SC and TH, roughly 10-20 optimization steps is sufficient to realize assembly. On the other hand, for more complex targets, like TTH and DIA, of order 100 iterations is required. In addition, each iterative step can vary in computational demand, depending on the nucleation kinetics. The cheapest was SC which used $6\times 10^{5}$ MD steps starting from a disordered configuration. Next in terms of computational demand was the TH crystal; this used $8\times 10^{6}$ total MD steps with half of the steps devoted to a slow, linear cooling back to the optimization temperature ($T$) from a hot initialization ($1.5T$). Finally, for the TTH and DIA crystals $1.5\times 10^{7}$ simulation steps were used to incorporate (1) a quick heat and slow cool initialization and (2) a sufficiently long equilibration so as to allow for the crystal to propagate across the long dimension from one seed face to another (periodic image). Specifically, for TTH and D we used corresponding temperature points $[5.0T, 1.8T, T, T]$ at corresponding MD steps $[0, 3\times 10^{5}, 1\times 10^{7}, 1.5\times 10^{7}]$ with linear temperature changes in between. Post-optimization, long equilibration runs (of order $1\times 10^{7}$ MD steps) for all crystals were performed to calculate $g(r)$ and extract configuration data for snapshots.

To study the effect of the seed in the TTH calculations, we adopted a combined equilibration and cooling protocol to generate the unseeded assemblies. First we discovered the temperature range in which the optimized interaction crystallized in the absence of a seed. The range for crystallization was empirically found to approximately lie between $0.73T$ and $0.81T$ based off of a sharp change in the potential energy trace with temperature. Using this estimate, we performed a combined heat, assemble and equilibrate cycle with linear temperature changes between the sequential temperatures $[5.0T, 0.81T, 0.73T, 0.73T]$ and corresponding simulation steps $[0, 3\times 10^{5}, 1.5\times 10^{7}, 6\times 10^{7}]$. This was performed five times to yield a small ensemble of representative configurations. Each was then slowly annealed according to the specified temperatures, $[0.73T, 0.0001T, 0.0001T]$, and molecular dynamics steps $[0, 5.9\times 10^{7}, 6\times 10^{7}]$ with potential energy data sampled from the final $1\times 10^{6}$ steps. This final annealing protocol was also adopted for calculating the ground state energy of the seeded crystalline assembly after removal of the harmonic seed constraints.

Similar calculations were performed for the optimized DIA potential in the absence of a seed. Unlike TTH, the DIA potential did not exhibit a well-defined crystallization temperature but rather forms amorphous glass states, and therefore we modified the above annealing protocol to achieve a slow cooling over the large temperature gap that encompasses the glassy transition. First, we performed a cycle with linear temperature changes between the following sequential temperatures $[5.0T, 1.25T, 0.71T]$ and corresponding simulation steps $[0, 3\times 10^{5}, 6\times 10^{7}]$. Five separate configurations were generated and each slowly annealed according to the specified temperatures, $[0.71T, 0.0001T, 0.0001T]$, and molecular dynamics steps $[0, 5.9\times 10^{7}, 6\times 10^{7}]$ with potential energy data sampled from the final $1\times 10^{6}$ steps. This final protocol was also used to approximate the ground state energy of the seeded DIA crystalline assembly after removal of the harmonic seed constraints.


%


\end{document}